\documentclass[%
 reprint,
 amsmath,amssymb,
 aps,
]{revtex4-2}

\usepackage{graphicx}
\usepackage{dcolumn}
\usepackage{bm}
\usepackage{amsmath}
\usepackage{mathtools}

\begin{document}

\preprint{APS/123-QED}

\title{TiO$_2$ doping effect on reflective coating mechanical loss for gravitational wave detection at low temperature}

\author{Yukino Mori} 
\author{Yota Nakayama}
\author{Kazuhiro Yamamoto}
\email{yamamoto@sci.u-toyama.ac.jp}
 \affiliation{Department of Physics, Faculty of Science, University of Toyama, Toyama, Toyama, Japan, 930-8555}
\author{Takafumi Ushiba}
 \affiliation{%
 Institute for Cosmic Ray Research, KAGRA Observatory, The University of Tokyo, Kamioka-cho, Hida City, Gifu 506-1205, Japan\\
}
\author{Danièle Forest}
\author{Christophe Michel}
\author{Laurent Pinard}
\author{Julien Teillon}
 \affiliation{Université Lyon, Université Claude Bernard Lyon 1, CNRS, Laboratoire des Matériaux Avancés
(LMA), IP2I Lyon / IN2P3, UMR 5822, F-69622 Villeurbanne, France}
\author{Gianpietro Cagnoli}
 \affiliation{Université de Lyon, Université Claude Bernard Lyon 1, CNRS, Institut Lumière Matière, F-69622
Villeurbanne, France}




\date{\today}

\begin{abstract}
We measured the mechanical loss of a dielectric multilayer reflective coating (ion-beam-sputtered SiO$_2$ and Ta$_2$O$_5$) with and without TiO$_2$ on sapphire disks between 6 and 77 K. The measured loss angle exhibited a temperature dependence, and the local maximum was found at approximately 20~K. This maximum was $7.0 \times 10^{-4}$ (with TiO$_2$) and $7.7 \times 10^{-4}$ (without TiO$_2$), although the previous measurement for the coating on sapphire disks showed almost no temperature dependence (Phys. Rev. D 74 022002 (2006)).  We evaluated the coating thermal noise in KAGRA and discussed future investigation strategies.
\end{abstract}

\maketitle


\section{Introduction}

Gravitational waves are ripples of space-time generated by the acceleration of masses, and predicted by the theory of general relativity on 1916. The LIGO-Virgo collaboration detected a gravitational wave for the first time on September the 14th 2015~\cite{GW150914} using two LIGO interferometers. This is one of the main scientific achievements of this century, and is considered the birth of gravitational wave astronomy. 
In the three observation runs that followed from 2015 to 2020, the LIGO and Virgo interferometers detected 90 events~\cite{GWTC3}.
To further increase the detection event number effectively and analyze the gravitational wave signals more precisely, noise reduction of the gravitational wave detectors is necessary, especially in the frequency band where the interferometers are the most sensitive, that is, approximately 100 Hz. 

In this frequency range, the dominant noise was the thermal noise of the mirror's internal elastic modes \cite{Saulson review}. This noise was dominated by the contribution of the reflective coating \cite{Levin 1998, Harry 2002}. This dielectric multilayer reflective coating has a high reflectance (at least 99.8\%)  and consists of Ta$_{2}$O$_5$ and SiO$_2$. LIGO and Virgo adopted TiO$_2$ doping to reduce the mechanical dissipation in the coating \cite{Granata 2020} because the coating thermal noise amplitude is proportional to the square root of the product of the dissipation and temperature \cite{Saulson review}. 

KAGRA \cite{KAGRA PTEP1} was the first kilometer-scale cryogenic underground gravitational wave detector. These two unique key features (cryogenic and underground) result in the reduction in thermal and seismic noise. The KAGRA main mirrors were cooled (to approximately 20~K). A previous study showed that cooling effectively suppresses the coating thermal noise \cite{Yamamoto 2006, Hirose 2014, Hirose 2020}. The coatings used in these studies and the KAGRA mirror had no TiO$_2$ doping. However, for KAGRA itself as well as future detectors, such as the Einstein Telescope \cite{ET review, ET design document}, further reduction is necessary and investigations are actively in progress \cite{Granata 2020, Granata 2016, Reid 2016, Steinlechner 2017}.

We investigated the TiO$_2$ effect at cryogenic temperature. The measurement of the coating (SiO$_{2}$/Ta$_2$O$_5$) with TiO$_2$ on Si has been reported \cite{Granata 2013}. We measured the coatings with and without TiO$_2$ on sapphire, which is the same material as the KAGRA main mirrors. From our measurement result, we derived the thermal noise of the KAGRA coating with and without TiO$_2$-doping.

\section{Experiment}

\subsection {outline}
Because the coating is extremely thin (of the order of $\mu$m), it is difficult to manufacture only the coating. We measured the loss angle of the sapphire disks with ($\phi_{\mathrm{with \,coating}}$) and without ($\phi_{\mathrm{without\,coating}}$) coating, and derived the coating mechanical loss angle ($\phi_{\mathrm{coating}}$) from their difference as follows: 
\begin{equation}
\phi_{\mathrm{coating}}=\frac{E_{\mathrm{sapphire}}}{E_{\mathrm{coating}}}(\phi_{\mathrm{with \,coating}}-\phi_{\mathrm{without\,coating}}),
\label{eq:coating loss}
\end{equation}
where $E_{\mathrm{sapphire}}$ and ${E_{\mathrm{coating}}}$ are the elastic energies of the sapphire disk and the coating, respectively. This energy ratio is of the same order as the ratio of the thickness of the sapphire disk to the coating and includes the difference in the elastic modulus.

In this investigation, we adopted the ringdown method to measure the sapphire disk loss. The resonant vibration of the disk was excited, and the decay resonant motion was measured. The displacement of the sapphire disk $x(t)$ is proportional to $\exp(-\pi f_0  \phi t) \sin (2 \pi f_0 t)$, where $f_0$ is the resonant frequency. The value $\phi$ is $\phi_{\mathrm{with\ coating}}$ or $\phi_{{\mathrm{without\ coating}}}$. Note that $\phi_{\mathrm{with\ coating}}$ and $\phi_{{\mathrm{without\ coating}}}$ are the inverse numbers of the Q values of the sapphire disk resonant mode with and without coating, respectively.

\subsection {Samples}

\begin{table}[b]
\caption{\label{tab:sample}%
Specifications of sapphire disk samples. The sapphire disks with coating (sample 1 and 2) were annealed (500~Celsius degree, 10~hours) as the mirrors of gravitational wave detectors \cite{Granata 2020}.
}
\begin{ruledtabular}
\begin{tabular}{cccc}
\textrm{}&
\textrm{Coating}&
\textrm{Layer numbers}&
\textrm{Coating thickness}\\
\colrule
1 & TiO$_2$ doping & 38 & 5.91~$\mu$m\\ 
2 & no TiO$_2$ doping & 40 & 6.41~$\mu$m\\ 
3 & no coating & 0 & 0~$\mu$m\\
\end{tabular}
\end{ruledtabular}
\end{table}

\begin{figure}[b]
\includegraphics{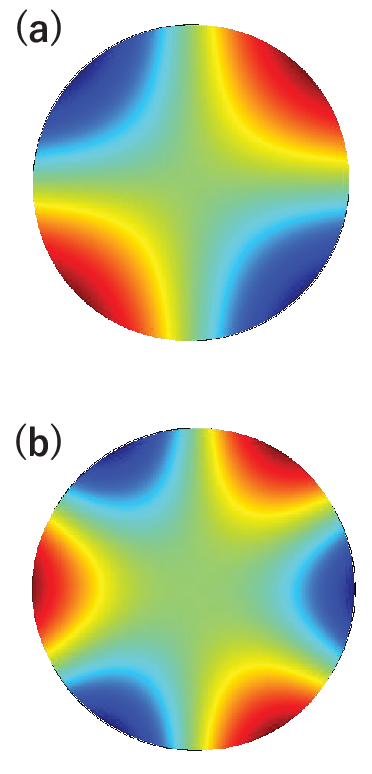}
\caption{\label{fig:modeshape} Shapes of the 1st (a) and 3rd (b) modes of the sapphire disk. Their frequencies are 540~Hz and 1.2~kHz.}
\end{figure}

The reflective coating used for gravitational wave detection was an ion-beam-sputtered amorphous dielectric multilayer. The layers of Ta$_2$O$_5$ and SiO$_2$ are stacked sequentially. LIGO and Virgo adopted TiO$_2$-doped Ta$_2$O$_5$ instead of pure Ta$_2$O$_5$ because TiO$_2$ doping reduces mechanical loss at room temperature \cite{Granata 2020}. We prepared reflective coatings with and without TiO$_2$ doping on the sapphire disks. Both coatings were deposited at LMA (Laboratoire des Matériaux Avancés)-Lyon in France, where all main mirrors coatings for LIGO, Virgo, and KAGRA were deposited. The coating with TiO$_2$ was used as a LIGO mirror witness; the sapphire disk and LIGO mirrors were coated simultaneously. The mixing ratio was Ti/Ta=0.27 \cite{Granata 2020}. The coating without TiO$_2$ was the same as that of the KAGRA mirrors. 

The diameter of the sapphire disk used as the substrate was 100 mm and its thickness was 0.5 mm. The c-axis was perpendicular to the flat surface. They were provided by SHINKOSHA \cite{Shinkosha} as per our previous measurement \cite{Yamamoto 2006}. The diameter of the coating on the sapphire disk was 90 mm. Table~\ref{tab:sample} summarizes the sample details. As a reference, a blank (no coating) sapphire disk with the same dimensions was prepared. 

We measured the resonant modes $\phi_{\mathrm{with\ coating}}$ and $\phi_{{\mathrm{without\ coating}}}$ at 540 Hz and 1.2 kHz, which were close to the frequency of the gravitational waves to be detected. These are the 1st and 3rd resonant modes. Figure~\ref{fig:modeshape} shows the shapes of these vibration modes obtained using COMSOL \cite{COMSOL}, which is a software program that performs calculations using the finite element method.

\subsection {Experimental equipment}

\begin{figure}[b]
\includegraphics[width=96mm]{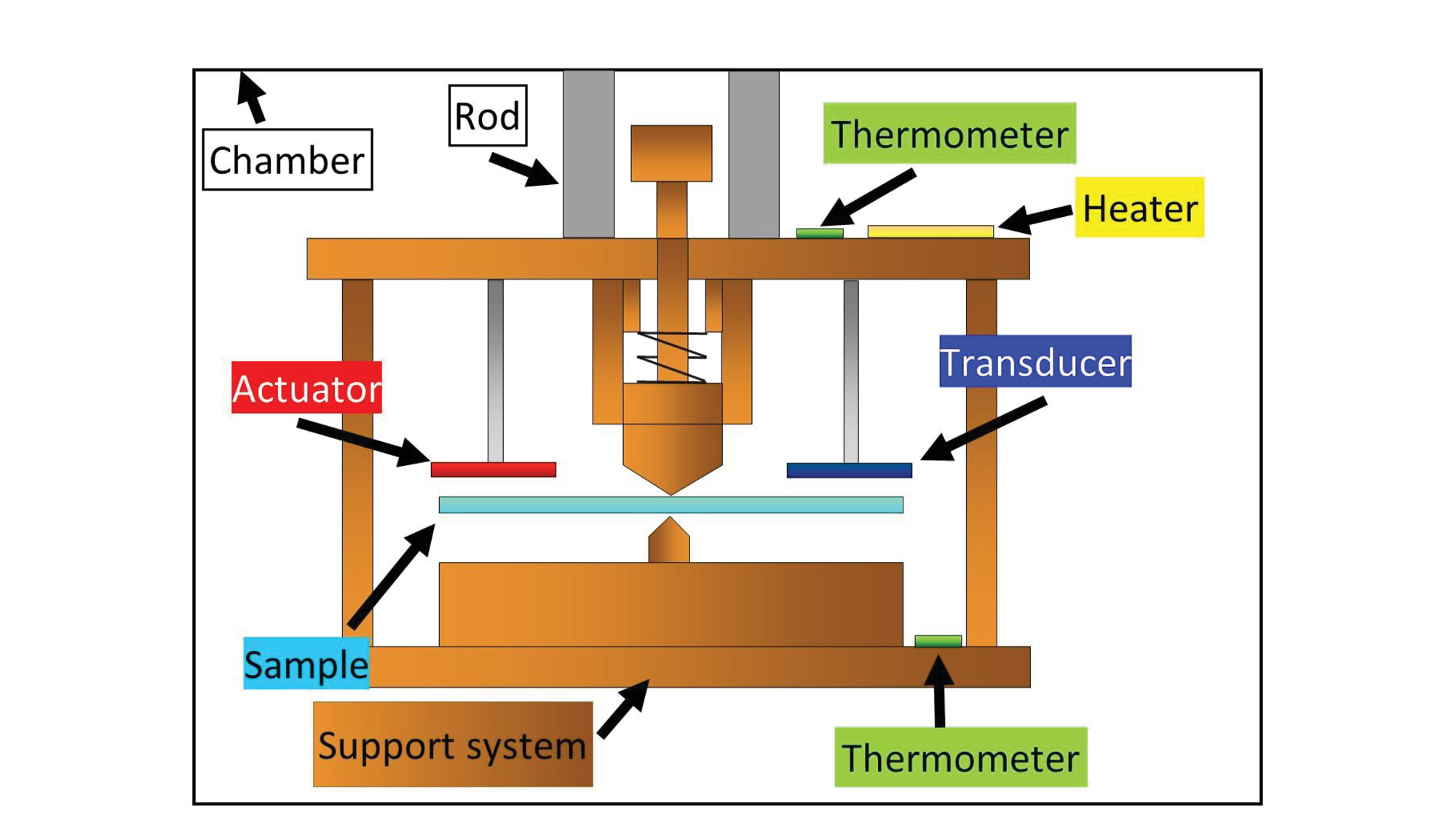}
\caption{\label{fig:chamber} Schematic side view of the apparatus in our vacuum chamber. This system is almost same as our old apparatus in Ref.~\cite{Yamamoto 2006,Mori 2023}. Only the center (nodal point) of the sapphire disk ("Sample" in this figure) was fixed in the nodal support system \cite{Numata 2000} made of copper. An electrostatic actuator to excite the resonant vibration was installed. An electrostatic transducer monitored the decay of the resonant motion. The stainless-steel rods ("Rod" in this figure) connected the top of the nodal support system to a vacuum chamber. This chamber was in the Dewar. The liquid nitrogen or helium was introduced into this Dewar. The pressure in the cooled chamber was approximately $10^{-3}$ Pa. For temperature control, a film heater was fixed on the top of the nodal support system. The two thermometers showed the temperature at the top and the bottom of the nodal support system.} 
\end{figure}

Our experimental equipment is similar to that used in the first measurement of coating mechanical loss at cryogenic temperatures~\cite{Yamamoto 2006,Mori 2023}. A vacuum chamber containing the experimental apparatus and sapphire disk sample (Fig.~\ref{fig:chamber}) were placed inside Dewar. Liquid nitrogen and helium were introduced to cool the chamber, apparatus, and the sapphire disk. The pressure in the cooled chamber was approximately $10^{-3}$ Pa. 
The vacuum chamber and apparatus were connected using stainless-steel rods with low thermal conductivity. During the initial cooling, gas was introduced to cool the apparatus and the sapphire disk. The chamber was evacuated because the residual gas caused additional loss to the sapphire disk. 

The support of the sapphire disk must be carefully designed because the vibration of the support itself caused by the resonant motion of the sapphire disk can add loss to the sapphire disk. To reduce this loss, a nodal support system \cite{Numata 2000} was adopted in which only the center of the sapphire disk was fixed. The disk center is the node of the measured 1st and 3rd modes. 
The diameter of the contact area between the nodal support system and sapphire disks was 2 mm. An electrostatic actuator was used to excite the resonant elastic mode. The displacement of elastic vibration was observed using an electrostatic transducer. Although dissipation was introduced by the transducer, we confirmed that it was sufficiently small (the measured sapphire disk dissipation was independent of the bias voltage applied to the transducer). 

After cooling with liquid nitrogen and helium, the temperature was adjusted using a heater attached to the top of the nodal support system. It was difficult to attach a thermometer to the sapphire disk because of dissipation by the thermometer itself. Therefore, the thermometers were attached to the top and bottom of the nodal support system. Because the nodal support system was made of copper and the heater was near the stainless-steel rod, it was expected that the temperature difference between the upper and lower parts of the nodal support system and the sapphire disk would be small. To confirm this hypothesis, a thermometer was attached to a sapphire disk without coating, and the temperatures of the sapphire disk and nodal support system were measured. The temperature difference was less than 0.5 K in the steady state (the deviation from the standard thermometer calibration curve was approximately between 0.25 K and 1 K). This sapphire disk with the thermometer was not used as the reference disk for the loss measurement.

\section{Results}

\begin{figure}[b]
\begin{minipage}{8.6cm}
\includegraphics[width=86mm]{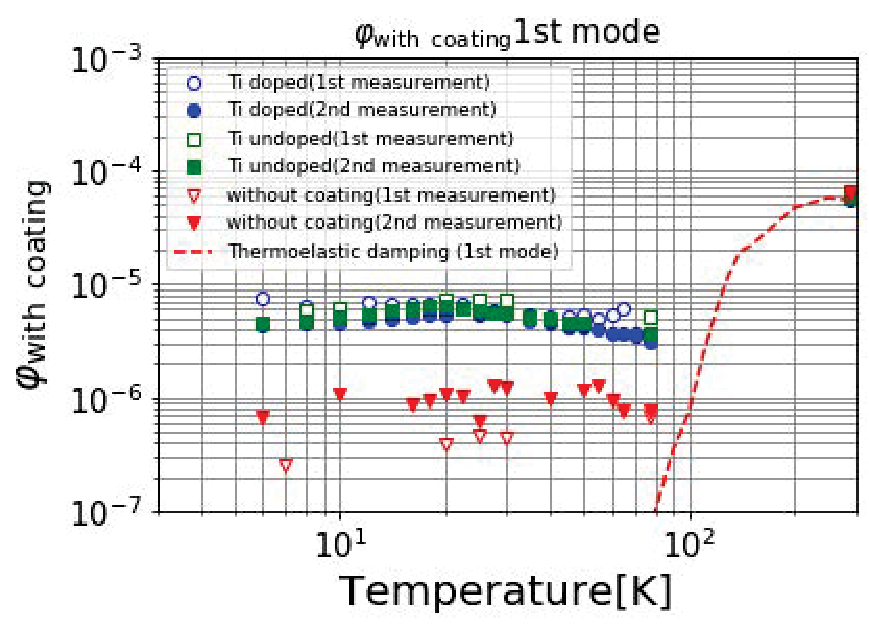}
\end{minipage}
\quad
\begin{minipage}{8.6cm}
\includegraphics[width=86mm]{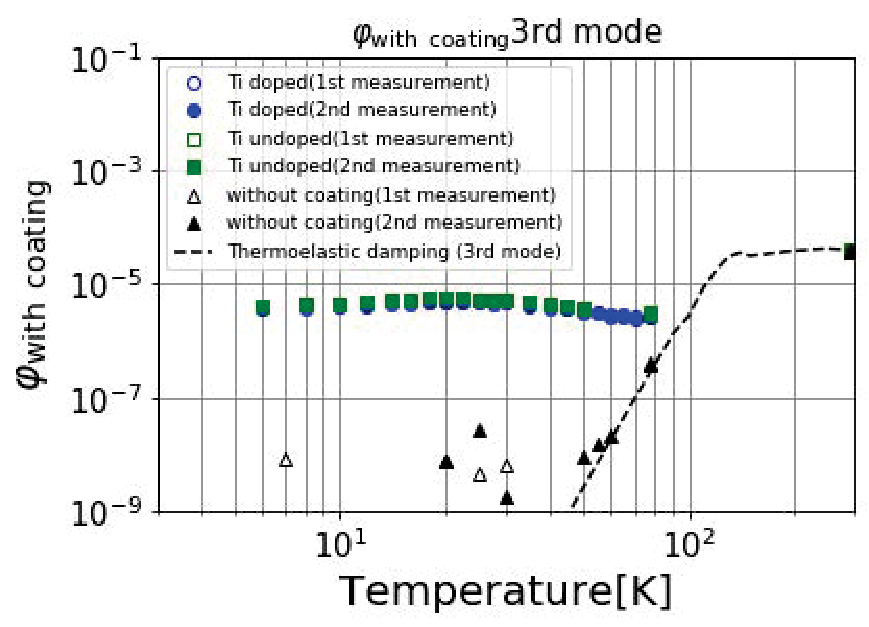}
\end{minipage}
\caption{\label{fig:measured phi} {Measured loss angle ($\phi$) of the sapphire disks with and without the coating. The circles and squares (blue and green in online) markers are loss angle of sapphire disks with TiO$_2$ doped and undoped coating, respectively. The triangles (red in online) markers are loss angles of sapphire disks without coating. Between the 1st (open markers) and 2nd (closed ones) measurements, the chamber was opened and the disk was removed and reinstalled. The dashed lines represent the loss angle limited by the thermo-elastic damping \cite{Blair 1982}. }}
\end{figure}

\begin{table}
\caption{\label{tab:COMSOL}%
Ratio of elastic energy in substrate to that in coating, $E_{\mathrm{sapphire}}/E_{\mathrm{coating}}$, calculated by COMSOL.
}
\begin{ruledtabular}
\begin{tabular}{lcc}
\textrm{}&
\textrm{1st mode}&
\textrm{3rd mode}\\
\colrule
TiO$_2$ doped & 126 & 143 \\
TiO$_2$ undoped & 119 & 135 \\
\end{tabular}
\end{ruledtabular}
\end{table}

\begin{figure}
\begin{minipage}{8.6cm}
\includegraphics[width=86mm]{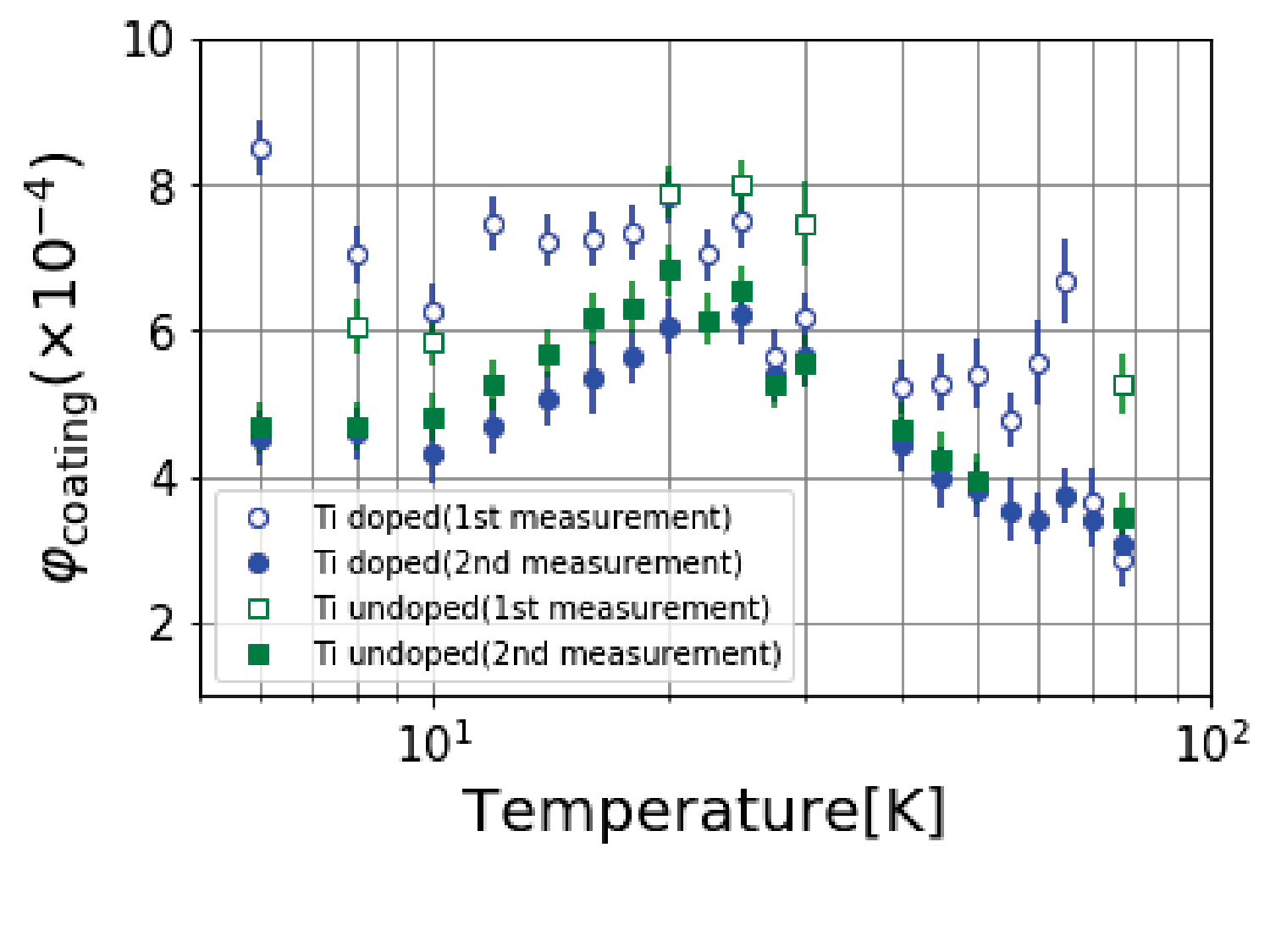}
\end{minipage}
\quad
\begin{minipage}{8.6cm}
\includegraphics[width=86mm]{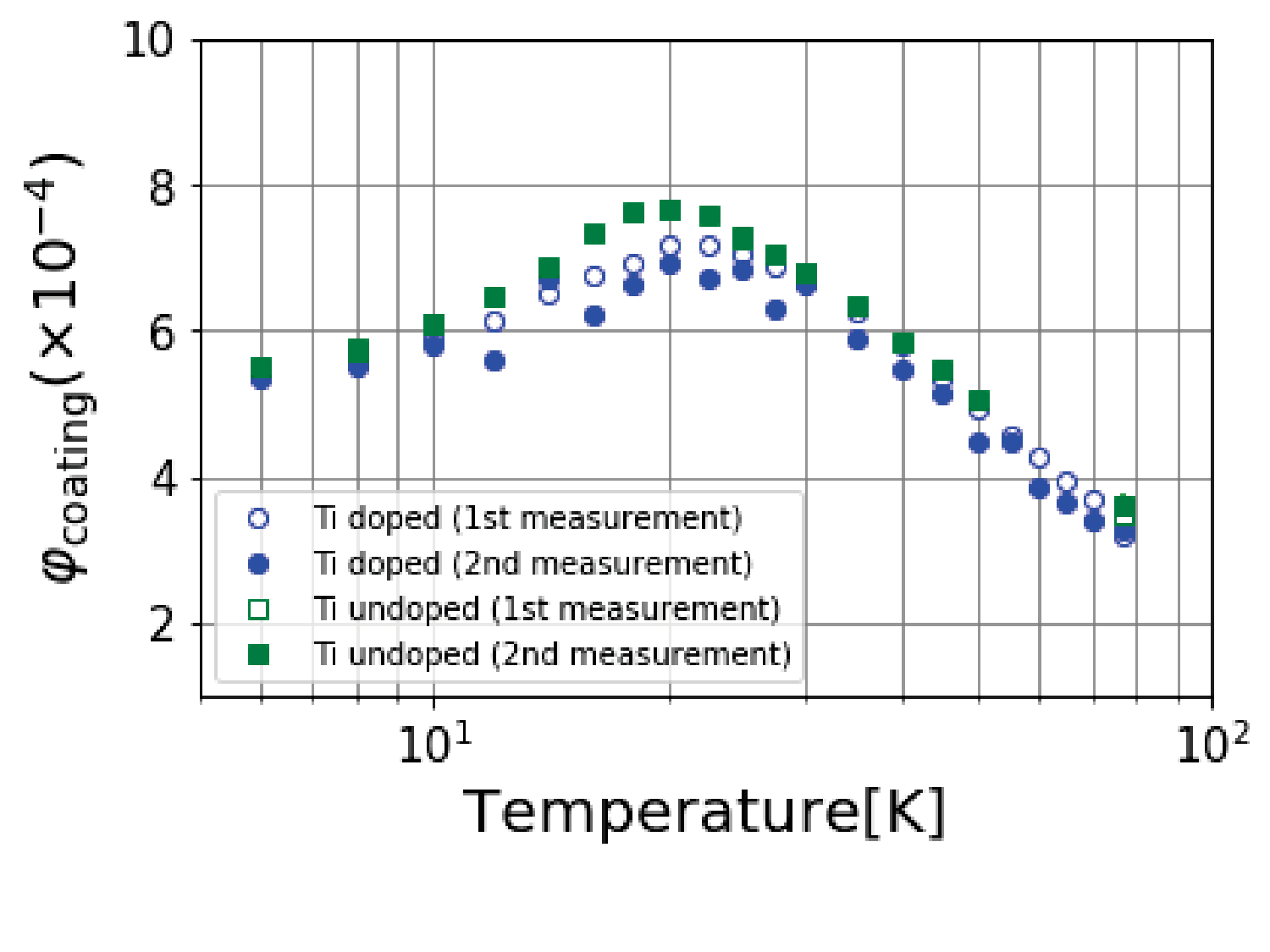}
\end{minipage}
\caption{\label{fig:coating} {Measured mechanical loss angles of the coating as a function of the temperature. The circles and squares (blue and green in online) markers represent the coating loss angles with and without TiO$_2$ doping, respectively. Between the first (open markers) and second (closed ones) measurements, the chamber was opened and the disk was removed and reinstalled. Note that the first measurement (without TiO$_2$) was at only 8~K, 10~K, 20~K, 25~K, 30~K, and 77~K. The difference between the first and second measurement of the 3rd mode (without TiO$_2$) is considerably small to observe this graph. The error bars of the 3rd mode were smaller than the size of the makers.}}
\end{figure}

Although the nodal support system was designed to fix the exact center of the sapphire disk, a discrepancy between the contact and disk centers could occur. In addition, this discrepancy changed in each measurement and reduced reproducibility. To evaluate this effect, we repeated the measurement twice as follows: after measurement at cryogenic temperatures, the chamber was opened, and the disk was removed. Thereafter, the same disk was set, and loss at cryogenic temperature was measured. We call “1st” and “2nd” measurements before and after the reset of the sapphire disk. The procedure for the one-cycle measurements is as follows: First, the disks were cooled in liquid nitrogen. The loss was measured at the nitrogen temperature. The disk was cooled using liquid helium. After cooling, the heater was switched on to adjust the temperature. The loss was measured between 6 K and 70 K. At each temperature, measurements were performed three or four times in both the 1st and 3rd modes with coating. 

The measured losses for each mode are shown in Figure \ref{fig:measured phi}. The average values (markers) and standard deviations (error bars) of three measurements are shown. The losses of the sapphire disk without the coating were approximately $1 \times 10^{-6}$ (1st mode) and $1 \times 10^{-8}$ (3rd mode). The loss of the 1st mode was larger because the vibration displacement around the disk center of the 1st mode was larger, and the loss by the nodal support system was effective. However, the loss of the sapphire disk with the coating was around $5.0 \times 10^{-6}$. The effect of the coating loss was clearly evaluated. 

The coating loss angles were derived from the values in Fig. \ref{fig:measured phi} and Eq.~(\ref{eq:coating loss}). The ratio $E_{\mathrm{sapphire}}/E_{\mathrm{coating}}$ was calculated using COMSOL, and is summarized in Table \ref{tab:COMSOL}.  
Figure \ref{fig:coating} shows the mechanical loss angle of the coating derived from the measurements. The error bar indicates the standard deviation of the measurements of the disk with and without coating. From Fig.\ref{fig:coating}, it can be confirmed that both the 1st and 3rd modes have a peak at approximately 20 K. The losses in the 3rd mode in the 1st and 2nd measurement were comparable. However, the coating loss of the 1st mode in the 1st measurement is greater than that in the 2nd measurement. This could be because the sapphire disk might have been moved from the center by some impact, and the effect of the nodal support system changed. The results of the 3rd mode are discussed in the following section.

\section{Discussion}

\subsection{Temperature dependence}

We found a temperature dependence (peak around 20~K) of SiO$_2$/Ta$_2$O$_5$ coating mechanical loss with and without TiO$_2$ doping. References \cite{Martin 2009} and \cite{Granata 2013} also reported peaks at approximately 20~K for Ta$_2$O$_5$ with and without TiO$_2$ doping and SiO$_2$/Ta$_2$O$_5$ loss with TiO$_2$ doping, respectively. In both studies, the coating was deposited on a silicon substrate. The latter was deposited by LMA-Lyon. 

According to previous measurements in Japan \cite{Yamamoto 2006, Hirose 2014}, no peak was found (the measured loss was almost independent of temperature) for SiO$_2$/Ta$_2$O$_5$ without TiO$_2$ doping on a sapphire disk. The coatings in these studies were prepared by the Japan Aviation Electronics Industry (which provided the coating for TAMA \cite{TAMA2001}) and the National Astronomical Observatory of Japan. We expect that the peak height depends on the details of the coating deposition process, and that the peak could be suppressed.

\subsection{TiO$_2$ doping effect}

Figure \ref{fig:coating} shows that TiO$_2$ doping slightly reduce the mechanical loss of the coating. The average 3rd mode loss angles of the 1st and 2nd measurement at 20~K (the target temperature of KAGRA) were $7.7 \times 10^{-4}$ (without TiO$_2$ doping) and $7.0 \times 10^{-4}$ (with TiO$_2$ doping). TiO$_2$ doping suppresses this loss by approximately 10\%. This corresponds to a 5\% reduction in the thermal noise amplitude. This reduction ratio is smaller than that at room temperature (25\% in loss angle \cite{Granata 2020}).

\subsection{Evaluation of KAGRA coating thermal noise}

\begin{figure}[b]
\includegraphics[width=86mm]{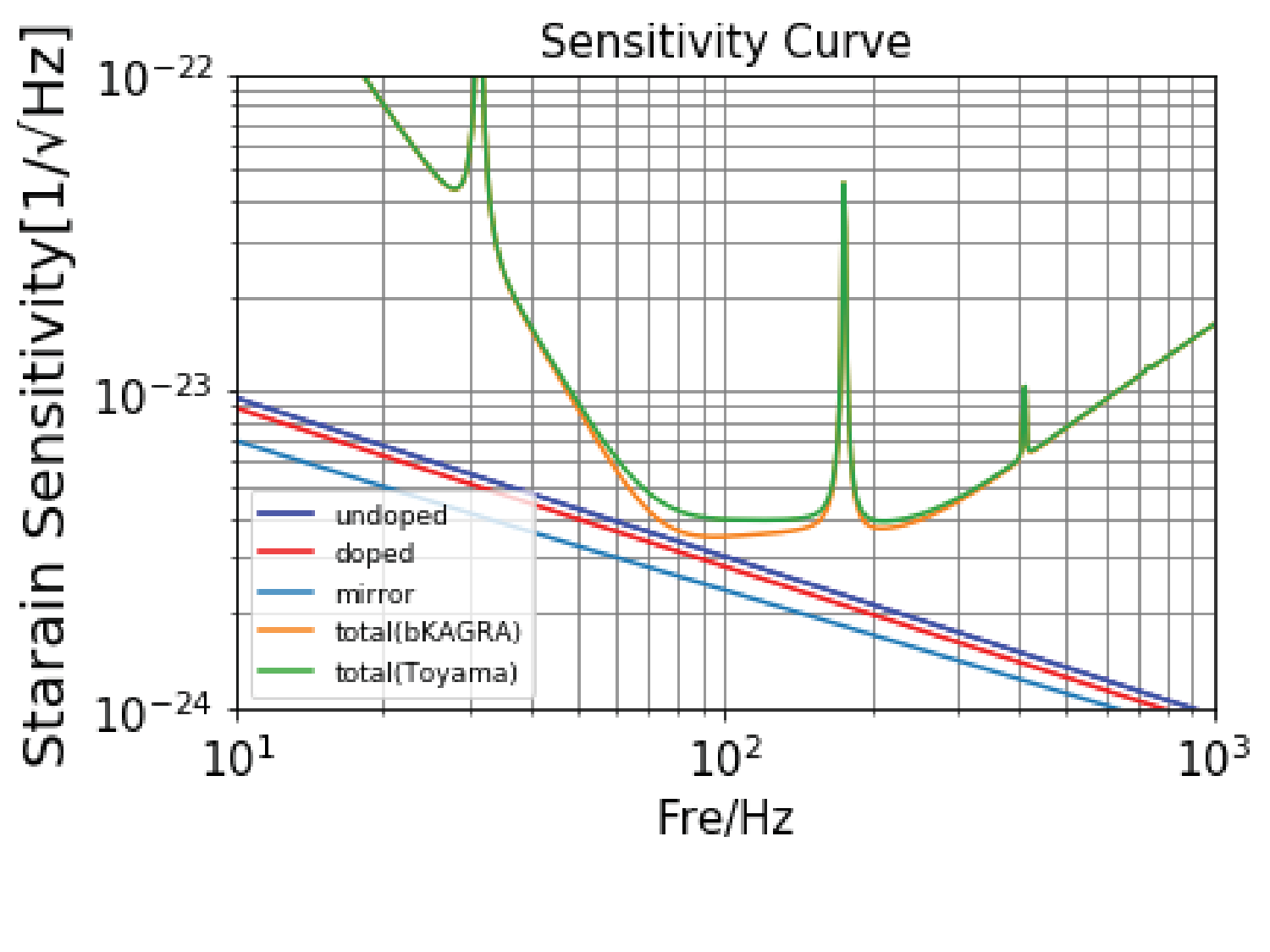}
\caption{\label{fig:KAGRA} Thermal noise of KAGRA. The temperature is 22~K, which is the operation temperature. The lines with "undoped" and "doped" represent the total thermal noise (including not only coating but also substrate) of mirror without and with TiO$_2$ derived from our measurement. The line with "mirror" is the total mirror thermal noise is derived from KAGRA official parameters \cite{KAGRA parameter 2017} which does not include our result. The "total(Toyama)" is KAGRA sensitivity, which includes coating thermal noise derived from our measurement result. The line with "total(bKAGRA)" is the KAGRA seneivity derived from KAGRA official parameter.}
\end{figure}

The thermal noise of the KAGRA coating was derived from our measurements as shown in Fig. \ref{fig:KAGRA}. This formula is introduced in Ref. \cite{Harry 2002}. The average loss angle of the 1st and 2nd measurements at 20 K for the 3rd mode were adopted. The mechanical loss of the coating was assumed to have no frequency dependence. The necessary parameters are summarized in Table. \ref{tab:KAGRA} \cite{KAGRA parameter 2017}. As reference, the mirror thermal noise and KAGRA sensitivity derived from KAGRA official parameter which does not include our result are shown.   

\begin{table}
\caption{\label{tab:KAGRA}%
KAGRA parameter for calculation coating thermal noise.
\cite{KAGRA parameter 2017}}
\begin{ruledtabular}
\begin{tabular}{lc}
\textrm{Properties}&
\textrm{Values}\\
\colrule
Baseline length &  3~km \\
Temperature & 22~K \\
Beam radius at the mirror & 35~mm \\
Wavelength of laser & 1064~nm \\
Number of coating layers for Input Test Mirror & 23 \\
Number of coating layer for End Test Mirror & 41 \\
Young's modulus of SiO$_2$ & 73.2~GPa \\
Young's modulus of Ta$_2$O$_5$ & 140~GPa \\
Young's modulus of sapphire & 400~GPa \\
Poisson ratio of SiO$_2$ & 0.164 \\
Poisson ratio of Ta$_2$O$_5$ & 0.23 \\
Poisson ratio of sapphire & 0.29 \\
Loss angle of sapphire & $10^{-8}$ \\
\end{tabular}
\end{ruledtabular}
\end{table}

The KAGRA mirror coating did not involve TiO$_2$ doping. At approximately 100~Hz in Fig. \ref{fig:KAGRA}, coating thermal noise contribution cannot be neglected. TiO$_2$ doping slightly reduces this thermal noise. The reduction method (deposition method, annealing, material with small loss, etc.) is investigated to improve the KAGRA sensitivity. This is useful even for future 3rd generation detector, such as the Einstein Telescope.

\section{Summary}

Although cooling is an effective method for reducing the thermal noise of interferometric gravitational wave detectors, further dielectric reflective coating mechanical loss reduction is required for the KAGRA and Einstein Telescope. The room-temperature detectors (LIGO and Virgo) adopted a TiO$_2$-doped coating because of its smaller loss. We measured the mechanical loss with and without TiO$_2$ doping on a sapphire substrate between 6 and 77 K, and evaluated the thermal noise of the KAGRA coating. 

The measured coating loss exhibited a temperature dependence with a peak at approximately 20 K. The peak values with and without TiO$_2$ doping were $7.0 \times 10^{-4}$ and $7.7 \times 10^{-4}$, respectively. A small reduction in TiO$_2$ was observed. Previous studies by other groups have shown similar results; nevertheless, they used silicon as a substrate. Previous investigations in Japan reported that coating loss was almost independent of temperature. We expect that this difference is owing to the details of coating deposition.

We evaluated the thermal noise of the KAGRA coating, and found that this contribution cannot be neglected. For the KAGRA and Einstein Telescope, further coating reduction investigations (smaller loss material, deposition methods, and annealing process) are necessary.

\section*{Acknowledgement}

We appreciate the valuable support of Masatake Ohashi, Takayuki Tomaru, Yoshiki Moriwaki, and Kanta Hattori.
The Engineering Machine Shop, Faculty of Engineering, University of Toyama, fabricated the parts of the nodal support system. The Low Temperature Quantum Science Facility at University of Toyama provided the liquid nitrogen and helium for this study. 
This work was supported by a JSPS Grant-in-Aid for Scientific Research(B) Grant Number 25287053, the First Bank of Toyama Scholarship Foundation, and the discretionary budget of the president of University of Toyama.


\end{document}